\documentclass[10pt,a4paper,twoside,twocolumn]{article}

\usepackage{blindtext} 

\usepackage[sc]{mathpazo} 
\usepackage[T1]{fontenc} 
\usepackage{microtype} 

\usepackage[english]{babel} 

\usepackage[hmarginratio=1:1,columnsep=20pt,margin=2.1cm]{geometry} 
\usepackage[hang, small,labelfont=bf,up,textfont=it,up]{caption} 
\usepackage{booktabs} 

\usepackage{lettrine} 

\usepackage{enumitem} 
\setlist[itemize]{noitemsep} 

\usepackage{abstract} 

\usepackage{titlesec} 
\renewcommand\thesection{\Roman{section}} 
\titleformat{\section}[block]{\large\scshape\centering}{\thesection.}{1em}{} 
\titleformat{\subsection}[block]{\large}{\thesubsection.}{1em}{} 

\usepackage{fancyhdr} 
\usepackage{titling} 

\usepackage{hyperref} 

\pretitle{\begin{center}\Huge\bfseries} 
\posttitle{\end{center}} 
\title{Slow-light analogue with a ladder of RLC circuits} 
\author{%
\textsc{J.-P. Cromi\`eres and T. Chaneli\`ere} \\[1ex] 
\normalsize Laboratoire Aim\'e Cotton, CNRS, Univ. Paris-Sud, ENS Cachan, Universit\'e Paris-Saclay, \\ 91405 Orsay Cedex, France \\ 
\normalsize \href{mailto:thierry.chaneliere@u-psud.fr}{thierry.chaneliere@u-psud.fr} 
}
\date{\today} 
\renewcommand{\maketitlehookd}{%
\begin{abstract}
\noindent
The linear susceptibility of an atomic sample is formally equivalent to the response of a RLC circuit. We use a ladder of lumped RLC circuits to observe an analogue of slow-light, a well-known phenomenon in atomic physics. We first characterize the radio-frequency response of the circuit in the spectral domain exhibiting a transparency window surrounded by two strongly absorptive lines. We then observe a delayed pulse whose group delay is comparable to the pulse duration corresponding to slow-light propagation. The large group delay is obtained by cascading in a ladder configuration doubly resonant RLC cells.
\end{abstract}
}

\usepackage{amsmath}
\usepackage{graphicx}
\usepackage{amsfonts}
\usepackage{amssymb}%
\newcommand{\x}{\ensuremath{\frac{Z_0}{Z_1}}}

\begin{document}

\maketitle


\section{Introduction}

Slow-light is a fascinating phenomenon that has been observed in a variety of atomic systems. When light passes through a medium with a large dispersion, its group velocity is greatly reduced. Light has been slowed to 17 m/s \cite{hau1999light} and even stopped during a minute \cite{heinze}. Despite its name, this phenomenon should not be restricted to the optical domain. Slow-light is due to a steep dispersion region where the group index is heavily modified as in strongly absorbing media in the vicinity of resonances. Atomic vapors in the optical domain are naturally well adapted because the absorbing lines are quite narrow and optically thick ensembles can be obtained. We propose to investigate these two conditions, narrow resonant lines and optically dense sample, by using a ladder of lumped RLC circuits.

A number of different electronic circuits have been proposed for displaying slow-light. The major pioneering experiments were performed by Nakanishi {\it et al.} \cite{nakanishi:1117,1193073,nakanishi:323}. Their approach used cascaded active low-pass filters and showed a very clear slow-light effect. Garrido Alzar {\it et al.} \cite{alzar:37} used a doubly resonant (coupled RLC) electronic circuit to simulate electromagnetically induced transparency (EIT), a slow-light precursor that has been extensively studied in multi-level atoms \cite{harris:36}. The work presented here is inspired by these previous experiments, but takes a different approach. First, we use passive RLC circuits as the elementary blocks in our slow-light system, in contrast to the active filters used by Nakanishi \cite{nakanishi:323}. Passive RLC circuits have the advantage to be formally equivalent to two-level atoms in their ground state as many of the slow-light experiments performed in atomic media.

More precisely, Garrido Alzar {\it et al.} \cite{alzar:37} discussed a coupled RLC circuits configuration to produce an EIT-like spectrum. This should produce a group delay in the time domain. We here propose to cascade elementary cells to increase the group delay. The idea is quite intuitive but has not been considered in details. In the main part of the paper, we study a ladder configuration on a factual basis and observe a net group delay. Different network topologies than a ladder could certainly be considered (bridged-T or lattice for example).

As a perspective in the appendix \ref{section_why}, we introduce a general representation derived from a transmission line model (telegrapher's equations). A description of a discrete series by a propagation equation has been already approached by Nakanishi \cite{nakanishi:323} with an array of amplifiers for slowly varying DC signal. The conditions of application have to be clarified thus defining the propagation constant, the forward and backward waves, the characteristic and input impedances of the line.

We propose to produce two closely spaced narrow resonances. As pointed out by Khurgin in his remarkable review \cite{Khurgin:10}, the so-called double resonance scheme can be understood easily and has been used very successfully to obverse slow-light in atomic medium \cite{camacho}. The paper is organized as follows. We first show that doubly resonant RLC cells can be cascaded in a ladder configuration to mimic the increasing absorption and dispersion of an extended atomic medium. We then realize the previous scheme with commercial lumped RLC components. The electronic circuit is characterized in the spectral domain. We finally observe a retarded pulse whose group delay is comparable to its duration. In the appendix, we discuss the application of a transmission line model for our configuration.

\section{A ladder of lumped RLC circuits}
\label{sec:ladder}

\subsection{A ladder of double resonance cells}\label{cell}

\begin{figure}
\centering
\fbox{\includegraphics[width=.99\linewidth]{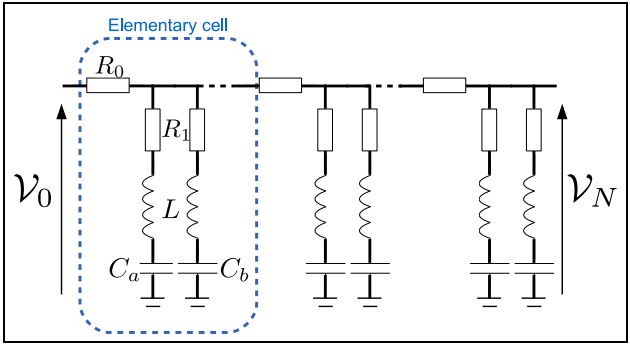}}
\caption{Ladder of double resonance cells. Each cell is composed of two series RLC circuits with ($R_1, L, C_a$) and ($R_1, L, C_b$) placed in parallel. The different cells are cascaded with a simple resistive coupling $R_0$.}
\label{fig:ladder}
\end{figure}

To obtain the double resonance scheme as described by Khurgin in \cite{Khurgin:10}, we propose to use two RLC circuits with slightly different resonance frequencies composing an elementary cell (see fig. \ref{fig:ladder}).

Two series RLC circuits with ($R_1$,$L$,$C_a$) and ($R_1$,$L$,$C_b$) are placed in parallel. They have different resonant frequencies $\omega_a=\displaystyle \frac{1}{\sqrt{L C_a}}$ and $\omega_b=\displaystyle \frac{1}{\sqrt{L C_b}}$ respectively but the same linewidth $\Gamma=\displaystyle \frac{R_1}{L}$. To carry on the analogy with the atomic vapor used by Camacho {\it et al.} in \cite{camacho}, our two RLC circuits would represent the D$_2$ hyperfine states of caesium. The resonances should be sufficiently separated to exhibit a transparency window between them leading to slow-light. In other words, each RLC resonator should have a high quality factor so that $\Gamma \ll \omega_a, \omega_b$. We can show that by cascading the number of elementary cells, the transmission mimics the behavior of a doubly resonant atomic medium.

\subsection{Lumped element modeling}\label{model}

The ladder circuit in fig. \ref{fig:ladder} can be modeled by using a transfer matrix for the cell numbered $n$ ($n$ ranging from $1$ to $N$ where $N$ is the total number of cells).

\begin{figure}[htbp]
\centering
\fbox{\includegraphics[width=.99\linewidth]{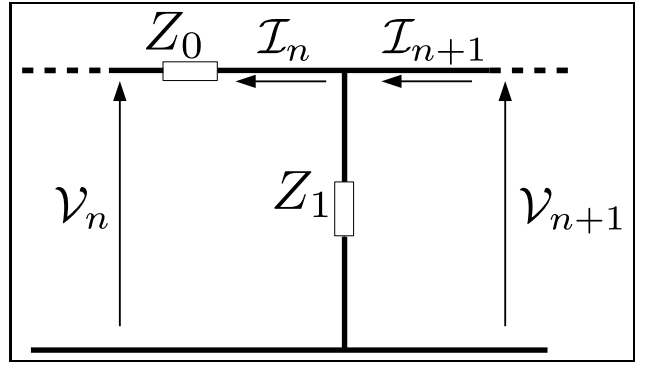}}
\caption{The input voltage and current of the cell number $n$ are $\mathcal{V}_{n}$ and $\mathcal{I}_{n}$ respectively. The output  voltage and current are then $\mathcal{V}_{n+1}$ and $\mathcal{I}_{n+1}$. $Z_0$ and $Z_1$ are generic complex impedances.}
\label{fig:bloc}
\end{figure}

The transfer matrix model in fig. \ref{fig:bloc} is general but in our case we have $Z_0=R_0$ and $Z_1(\omega)$ is the parallel impedance (written $/\!\!/$) of the two resonators $R_1, L, C_a$ and  $R_1, L, C_b$ namely

\begin{equation}\begin{split}
Z_1(\omega)=\displaystyle \left( R_1+jL\omega+\frac{1}{jC_a\omega}\right) \\ /\!\!/ \left( R_1+jL\omega+\frac{1}{jC_b\omega}\right) \label{Z1}
\end{split} \end{equation}
where $\omega$ is the signal frequency.

The output of $N$ cascaded cells can indeed be derived formally  from Kirchhoff's laws by writing the transfer matrix
\begin{eqnarray}\label{transfer_matrix}
A_{n}=\left[
\begin{array}{c}
\mathcal{V}_{n} \\
Z_0 \mathcal{I}_{n} \\
\end{array}\right] &=\left[
\begin{array}{cc}
1+\x & 1 \\
\x & 1 \\
\end{array}\right] \left[
\begin{array}{c}
\mathcal{V}_{n+1} \\
Z_0 \mathcal{I}_{n+1} \\
\end{array}\right]\\ &=M A_{n+1}
\end{eqnarray}
$A_n$ has the dimension of a voltage so the transfer matrix $M$ is dimensionless and is a function of $\omega$ that we omit for simplicity. Input and output of the circuit are related by the expression:
\begin{eqnarray}
A_{0}=M^N A_{N}
\end{eqnarray}

The exponential $M^N$ is calculated by diagonalizing $M=P D P^{-1}$ with
\begin{eqnarray}
P &=\left[
\begin{array}{cc}
\frac{Z_1}{Z_0} \Lambda_{+} & \frac{Z_1}{Z_0} \Lambda_{-} \\
1 & 1 \\
\end{array} \right]\\ & \mbox{\,\,\,  and   \,\,\,} \nonumber \\
D& =\left[
\begin{array}{cc}
1+\Lambda_{+}& 0 \\
0 & 1+\Lambda_{-} \\
\end{array}\right]
\end{eqnarray}
where \begin{equation} \Lambda_{\pm}=\displaystyle \frac{1}{2}\left({\x \pm \sqrt{\left(\x\right)^2+4\x}}\right) \label{Lambda} \end{equation}

The boundary conditions are imposed on one side by the applied input voltage $\mathcal{V}_{0}$ and on the other side by the high measurement load imposing $\mathcal{I}_N=0$.

The voltage transfer function $\displaystyle \frac{\mathcal{V}_{N}}{\mathcal{V}_{0}}$ is then given by the first coefficient of $M^N=P D^N P^{-1}$:

\begin{equation}\label{transfer_discrete}
{\mathcal{V}_{N}}= \mathcal{V}_{0}\frac{\Lambda_{+}+\Lambda_{-}}{\Lambda_{+}\left(1+\Lambda_{+}\right)^N - \Lambda_{-}\left(1+\Lambda_{-}\right)^N}
\end{equation}

The voltage transfer function depends on $\omega$. This defines the total transmission of the circuit {\it i.e.} its Bode diagram. It is complex with an amplitude (absorption) and a phase (dispersion). We will now show that the topology proposed in fig. \ref{fig:ladder} produces two closely spaced absorption peaks. Within the transparency window, the group delay increases as the peak absorption.

\subsection{Absorption spectrum and group delay}\label{abs_spectrum}

The expressions \eqref{Lambda} and \eqref{transfer_discrete} allow to calculate the transmission spectrum of the ladder in amplitude and phase. For the numerical application, we use the fitted values of $R_0, R_1, L, C_a$ and  $C_b$ that will be discussed later in \ref{exp_circuit}.

\begin{figure}[htbp]
\centering
\fbox{\includegraphics[width=.99\linewidth]{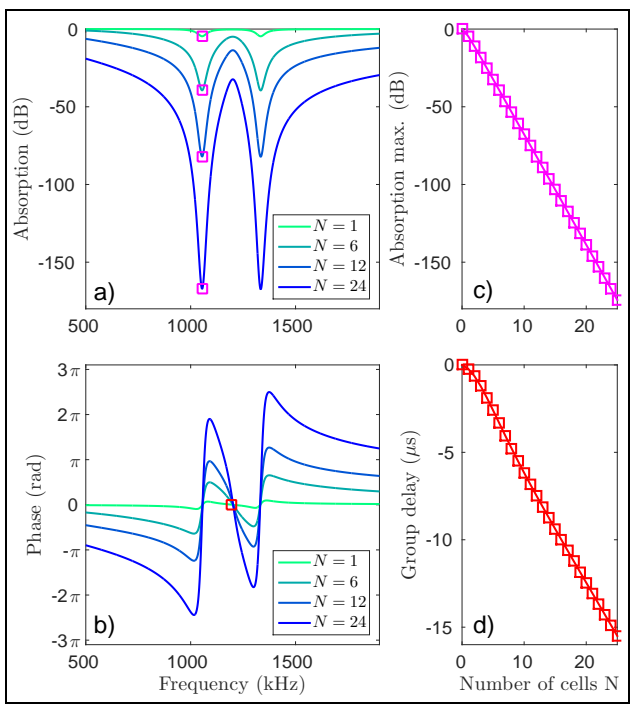}} 
\caption{Absorption (a) and phase (b) spectra of the RLC ladder for an increasing number of elementary cells $N=1,6,12$ and $24$. Peak absorption (c) corresponding to the square purple markers on the absorption spectrum for an increasing number of elements $N$. The group delay in the transparency window (red marker on the phase spectrum) is plotted as a function of $N$.}
\label{fig:Abs_delay_vs_Nelt}
\end{figure}

By plotting the absorption spectrum (fig.\ref{fig:Abs_delay_vs_Nelt}.a), we first observe the two resonances corresponding to the resonant excitation of $R_1, L, C_a$ and  $R_1, L, C_b$. The peak resonant absorption increases exponentially as the number of elementary cells (linear dependency in log scale of fig. \ref{fig:Abs_delay_vs_Nelt}.c). Cascading the elements in a ladder configuration mimics an atomic vapor whose absorption scales exponentially with the length of the medium (Bouguer-Beer-Lambert attenuation law).

Within the transparency window bounded by the two resonances, the slow-light effect can be evaluated by calculating the group delay. This later is deduced from the dispersion curve (fig. \ref{fig:Abs_delay_vs_Nelt}.b) by calculating the slope (first derivative) of the phase spectrum. The center of the window is marked by a red square in fig. \ref{fig:Abs_delay_vs_Nelt}.b. The group delay scales linearly with the number of cells $N$ essentially following the increasing peak absorption (fig. \ref{fig:Abs_delay_vs_Nelt}.c).

A ladder of double resonance cells resembles an atomic sample with two neighboring resonances. The resonant absorption coefficient goes linearly with the number of cells. Within the transparency window, as expected from the dispersion curve, the pulse propagation will be retarded by the group delay. This latter follows the resonant absorption as in an atomic vapor. The double resonance scheme \cite{Khurgin:10} has the advantage of the simplicity. Its main drawback comes from the imperfect absorption at the transparency window center (see fig. \ref{fig:Abs_delay_vs_Nelt}.a) because of the off-resonance excitation of the neighboring transitions. This is not the case for the EIT in which the total transparency is ensured by destructive interference in a $\Lambda$ level scheme  \cite{harris:36}. In the double resonance scheme, a trade-off should be found. A high resonant absorption is necessary to obtain a large group delay in the limit of the absorption of the slow-down pulse because of off-resonance excitation. This trade-off guides our experimental approach.

\section{Experiment}
\label{exp}
\subsection{Electronic circuit}
\label{exp_circuit}

We have designed our slow-light circuit  using commercially available components. We decided to implement our slow-light circuit for resonance frequencies in the MHz range. This choice is based on a number of practical considerations. First, the resistance for each element should be kept low to obtain narrow resonances, but it cannot be too low because otherwise too much current is drained out of the function generator. A few ohms per element, giving a few tens of ohms for the circuit, seems reasonable and compatible with the standard output impedance of a generator, typically 50 $\Omega$.

Assuming a few ohms for the resistance and a  MHz frequency will give an inductance in the $\mu$H range and a capacitance in the nF range. Thus, a quality factor larger than 10 is expected.  From a practical standpoint, this range is good to work in because there is a wide range of commercial inductors at $\mu$H and capacitors at nF. A nF capacitance is sufficiently high that it will dominate any parasitic capacitances, such as in the BNC cables. Meanwhile,  a $\mu$H inductance is sufficiently small that large DC and RF resistance (due to the winding skin effect), which can reduce the quality factor, are avoided. For our circuit we chose commercial inductors (Panasonic ELC08D270E) with a uniform inductance of $27 \mu$H and just 0.7\% RMS dispersity. This value is much lower than the commercial tolerance grade. It is important to verify that the inhomogeneity is low for inductors as they are known to be less precise than capacitors.

Based on these practical considerations, we chose $R_0=4.7 \Omega$, $R_1\simeq2.7 \Omega+4 \Omega$, $L=27 \mu $H, $C_a=680 $ pF and $C_b=1000 $ pF. The $R_1$ value includes the series resistance of the inductor ($\simeq 4 \Omega$), which we measured independently at 1 MHz. This is mostly due to the skin effect at this frequency.  For this set of values, the resonances are expected at $\omega_b=2\pi \times969$ kHz and $\omega_a=2\pi \times1174$ kHz.

\begin{figure}[htbp]
\centering
\fbox{\includegraphics[width=.99\linewidth]{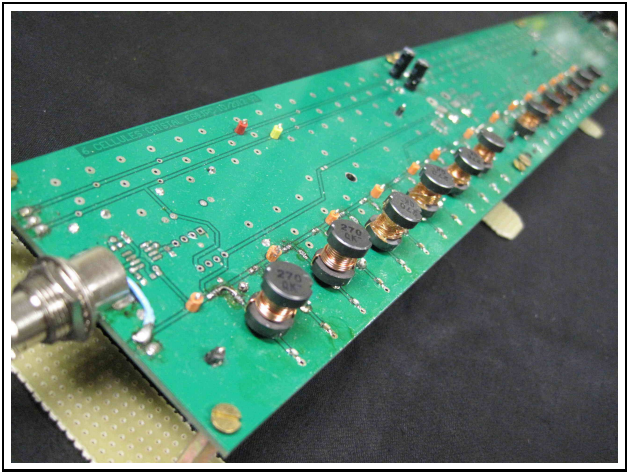}}  
\caption{Picture of a ladder of double resonance cells with $N=12$. Inductors are prominent as compared to the low profile of the surface-mount resistors and capacitors. 12 inductors are visible, the others are soldered on the back of the circuit. The distance between inductors has been chosen to avoid direct inductive coupling (top and the back rows are staggered).}
\label{fig:RLC_photo}
\end{figure}

With respect to the number of elements $N$, we chose 12 pairs of $C_a$ and $C_b$, meaning $N=12$ elementary cells in total. $N=12$ represents a good trade-off as simulated in fig. \ref{fig:Abs_delay_vs_Nelt}: a 7$\mu$s group delay and a 14 dB residual off-resonant absorption are expected. The soldering of 24 RLC elements circuit can be done without specific experience in electronics (see fig. \ref{fig:RLC_photo}). 

The circuit can now be characterized by measuring its transfer function (Bode diagram) and compared to the model developed in \ref{model}.

\subsection{Characterization of the spectral response}

We use a Tektronix  AFG3021B RF function generator as a tunable source to record the Bode diagram. A continuous sine wave was applied at the input of the circuit with a constant amplitude (typically $100$ mV). The output was recorded with a standard oscilloscope allowing to measure amplitude and phase.

\begin{figure}[htbp]
\centering
\includegraphics[width=.99\linewidth]{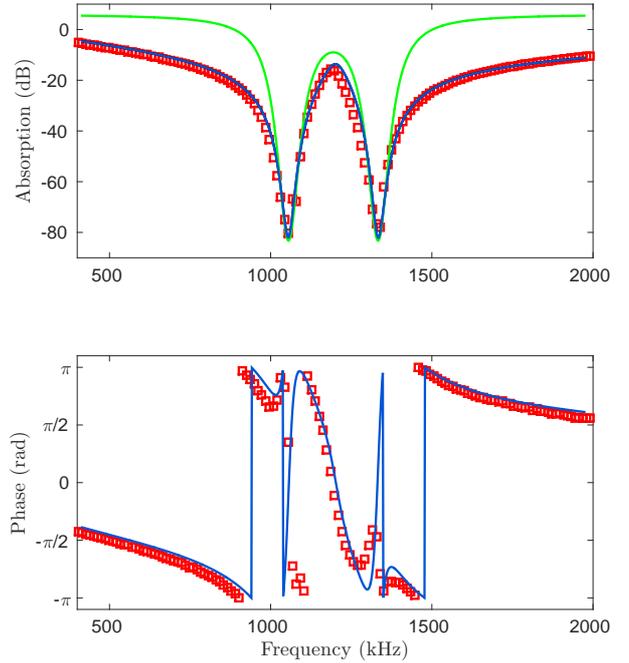}
\caption{Absorption (top) and phase (bottom) spectra of the RLC ladder for $N=12$ cells. The frequency was swept from 400kHz to 2MHz, covering the region of interest. The amplitude (top) and the phase (bottom) of the outgoing sine wave were measured (red squares). The blue lines represent the theoretical curves from  \eqref{Lambda} and \eqref{transfer_discrete} using the fitted values of $R_0^\mathrm{fit}, R_1^\mathrm{fit},L^\mathrm{fit}, C_a^\mathrm{fit}$ and  $C_b^\mathrm{fit}$ (see text). The green line (top) is deduced from the propagation equation discussed in appendix \ref{section_why} (eq. \ref{forward_gamma}).}
\label{Spectro_CW}
\end{figure}

In the transmission spectrum (fig.\ref{Spectro_CW}, top), we clearly observe two resonances at $\omega_a^\mathrm{fit}=2\pi \times 1055$ kHz and $\omega_b^\mathrm{fit}=2\pi \times 1330$ kHz. They are both deeply absorbing, 82dB. The half-width at half-maximum are typically $50$ kHz. We also plot the phase of the outgoing sine wave compared to the phase of the incoming one (fig.\ref{Spectro_CW}, bottom). Close to the resonances, the phase is extremely difficult to measure because of the high attenuation.

The measured resonances $\omega_a^\mathrm{fit}=2\pi \times 1055$ kHz and $\omega_b^\mathrm{fit}=2\pi \times 1330$ kHz deviates from the ones expected ($2\pi \times969$ kHz and $2\pi \times1174$ kHz) from the nominal values of the components. The deviation is most likely due to the parasitic capacitances of the circuit. In order to properly fit the experimental data, we leave $R_1^\mathrm{fit}$, $L^\mathrm{fit}$, $C_a^\mathrm{fit}$ and $C_b^\mathrm{fit}$ as free fitting parameters to additionally account for, for example, the internal resistance of the inductor and the additional inductance and capacitance introduced by the cables. The value of $R_0^\mathrm{fit}$ was fixed at its nominal value $4.7 \Omega$ (see \ref{exp_circuit}).  The solid blue lines in fig.\ref{Spectro_CW} are the result of this computation. We reproduce fairly well the experimental results with the parameters $R_1^\mathrm{fit}=6.7 \Omega$, $L^\mathrm{fit}=25 \mu $H, $C_a^\mathrm{fit}=570 $ pF and $C_b^\mathrm{fit}=910 $ pF, sufficiently close to the specification values. From these fitting parameters, we can extract a more accurate value of the linewidth, $\Gamma=\displaystyle \frac{R_1^\mathrm{fit}}{L^\mathrm{fit}}=2\pi \times 43$kHz, corresponding to Q-factors of 25 and 31 for the resonances $\omega_b^\mathrm{fit}=2\pi \times1055$ kHz and $\omega_a^\mathrm{fit}=2\pi \times1330$ kHz respectively.

Except close to the resonances  $\omega_b^\mathrm{fit}$ and $\omega_b^\mathrm{fit}$, the phase changes smoothly, following the curve predicted from eq.\eqref{transfer_discrete}. In particular,  within the transparency window, the phase changes quasi-linearly, and the slope directly gives the group delay. A steep variation of the phase (much larger than 2pi in our case, see fig.\ref{Spectro_CW}.b) is possible because of the accumulation through many cells. As a comparison, for a single RLC cell, the phase variation cannot be larger than 2$\pi$. This statement is intimately related to the slow-light effect because the group delay is precisely the first derivation of the phase variation.

Based on this data, we can expect to see slow-light in the time domain.

\subsection{Group delay measurement}

\begin{figure}[htbp]
\centering
\includegraphics[width=.99\linewidth]{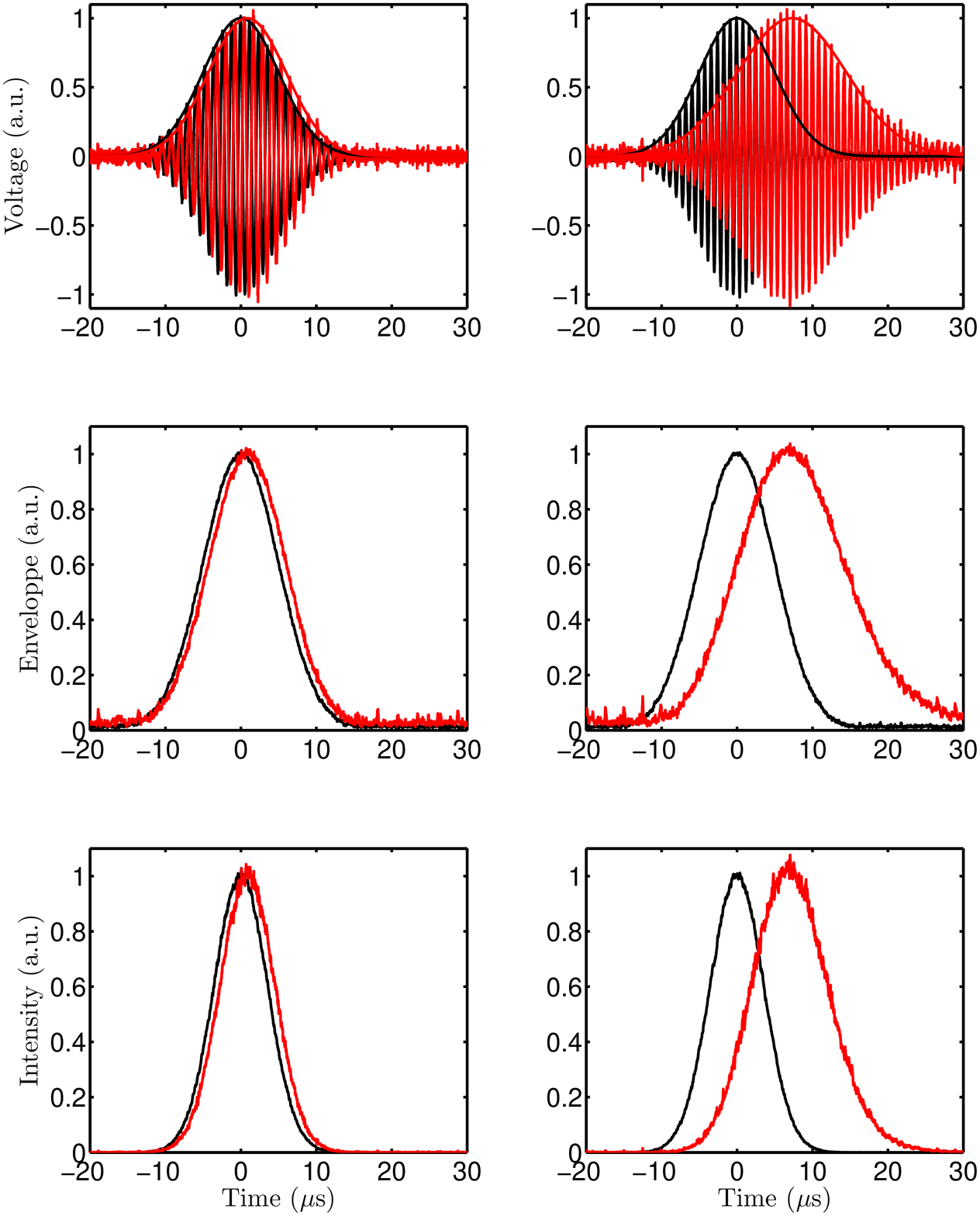}
\caption{Propagation of a $5\mu s$  Gaussian pulse (incoming in black and outgoing in red). Two different carrier frequencies are shown:  $840$ kHz in the left-hand column and $1188$ kHz in the right-hand column. In each column, the top line represents the normalized recorded data, the middle line is the pulse envelope (amplitude envelope of the analytic representation) and the bottom line is the intensity (square of the envelope). Instead of the incoming pulse, the equivalent free-space propagating pulse can sometime serve as reference. In our case, they are indistinguishable because our circuit board is 30cm long, so the free-space delay is typically a nanosecond well below the pulse duration.}
\label{Trait_SlowLightGenArb_article}
\end{figure}

To observe slow-light, we sent a short pulse into the circuit at the center of the transparency windows where the dispersion curve is linear. The pulse was chosen to be as short as possible but with a bandwidth that fit within the transparency window. Otherwise, pulse distortions would be expected. Longer pulses can be used as well but if they are too long compared to the measured delay, the visual impression is less demonstrative. So we chose a $5\mu s$ duration, corresponding to a 200kHz bandwidth. This filled approximately 2/3 of the $[\omega_a^\mathrm{fit} \,\,,\, \omega_b^\mathrm{fit}]$ window. The carrier frequency of the function generator was set to $2\pi \times 1188 $kHz $\simeq\, \frac{1}{2}(\omega_a^\mathrm{fit}+\omega_b^\mathrm{fit})$. The envelope was programmed using the arbitrary waveform mode of the function generator. The input and output pulses can be seen in black and red respectively in fig.\ref{Trait_SlowLightGenArb_article}.

As a reference, we represent in the left-hand column of fig.\ref{Trait_SlowLightGenArb_article} the pulse for a $840$ kHz carrier frequency (low dispersion). We use normalized curves because we are only interested in the delay measurement and not the attenuation. It should be kept in mind that the attenuation is far from negligible in the transparency window ($\sim$15dB), so the true amplitude at this frequency is much lower than in the reference case. 

At the center frequency $1188$ kHz (right column of fig.\ref{Trait_SlowLightGenArb_article}), the pulse is clearly delayed, by $7\mu s$ for the maximum of the pulse. The pulse is also distorted and elongated from its $5\mu s$ incoming duration to $7\mu s$. Distortions are expected because the incoming bandwidth is not negligible with respect to the transparency window. 

The group delay observed in the experiment is substantial, the same size as the pulse width, allowing the delay to be seen clearly. The delay is even more clearly observed in the intensity data (square of the amplitude envelope), as the intensity pulses are narrower by a factor of $\sqrt{2}$ than the amplitude pulses. Intensity pulses are a useful way of looking at the data, because this mimics the situation in an atomic system, for which  intensity measurements are much more straightforward to implement than field detection techniques.

\begin{figure}[htbp]
\centering
\includegraphics[width=.99\linewidth]{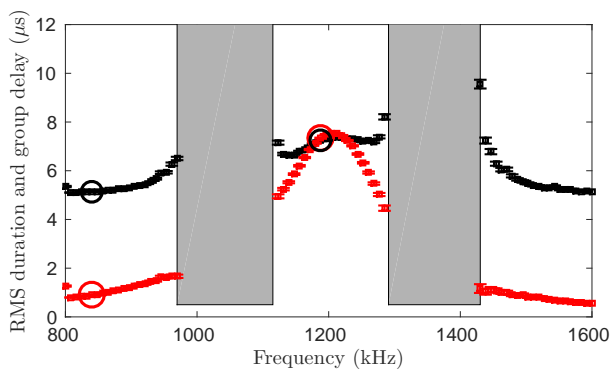}
\caption{Group delay measurements (in red) in the range 800kHz to 1600kHz. We also measure the RMS duration (in black) of the outgoing pulse envelope to characterize its elongation. When the attenuation is larger than 40dB (shaded areas), the fitting procedure is extremely inaccurate, yielding very large error bars, making the measurements irrelevant in this range so they have been discarded. The circles correspond to the frequencies of fig. \ref{Trait_SlowLightGenArb_article} ($840$ kHz and $1188$ kHz).}
\label{Trait_SeriesGenArb}
\end{figure}

These group delay measurements are reproduced for different carrier frequencies covering the range 800kHz - 1.6MHz in fig. \ref{Trait_SeriesGenArb} (red curve).
Dispersive measurements were extremely difficult when the signal was strongly attenuated ($> 40$ dB), regions which are represented by the two shaded areas in the figure. The fitting procedure gives a poor confidence interval: the measurements are irrelevant and have been discarded. The measurements are much more accurate in the transparency windows and off-resonance (both sides). 

The red trace in fig. \ref{Trait_SeriesGenArb} shows that off-resonance, for example at $840$ kHz (see fig.\ref{Trait_SlowLightGenArb_article}, left), the pulse exhibits a small delay because the slope dispersion profile is moderate (fig.\ref{Spectro_CW}, bottom).  In the transparency window, meanwhile, the dispersion curve is steeper yielding a substantial group delay $7\mu s$ as in fig.\ref{Trait_SlowLightGenArb_article} (right column). The group velocity peaks in the center of the transparency window, demonstrating clearly that it is a sum of the effects of the two resonators. 

We can finally evaluate the distortion of the outgoing pulse by computing its RMS duration (given by the fitting procedure) in fig.\ref{Trait_SeriesGenArb} (black curve). We retrieve the $5\mu s$ duration of the incoming pulse far off-resonance (no distortion), while substantial distortions are seen near the two resonances, as well as in the transparency window. As previously mentioned, distortions are expected in the transparency window because of the finite incoming bandwidth.

\subsection{Discussion}

In our case, the distortion results from the usual trade-off observed in optical slow-light and is independent of the physical system. As already discussed, for a given group delay, the incoming pulse should be made as short as possible to increase the delay-to-bandwidth product. This latter being a figure of merit for slow-light propagation. Distortion appears because the pulse bandwidth is not strictly contained in the transparency window. More precisely, the bandwidth cannot only be represented by a linear dependency of the phase (giving the group delay). The first higher order term is due to the absorption on the side of the transparency window. In other words, the wings of the pulse bandwidth are more absorbed than the center, as a consequence the pulse is elongated in the time domain.
As compared to the double resonance scheme in atomic medium that we keep as a point of comparison \cite{camacho}, the absorption (or losses) have a different but somehow comparable origin. In our case, the losses are due to the restrictive part of the circuit, meaning heat transfer in the resistors. This is not the case for an atomic vapor in which the decay term is due to photon scattering. This is interpreted as losses because the scattered photons leave the propagating mode under consideration. It should be nevertheless noted that the atomic decay terms could also represent non-radiative decays as atomic like impurities in solids for example leading to heat transfer. So the analogy between atoms and RLC circuits with non-radiative and resistive losses could be pushed on step further.

\section{Conclusion}

In summary, our setup produces a clearly observable slow-light effect in a ladder of RLC cells in a double resonance scheme. The observed delay in comparable to the pulse duration corresponding to a time-to-bandwidth product close to unity. We also show that cascading elementary cells increases the absorption and dispersion both being related by the Kramers-Kronig relation due to the causal character of the setup.

The cascaded elementary blocks provide an interesting resemblance with highly optically absorbing materials. The analogy can certainly be pushed one step further by comparing the structure of the Maxwell equations in an atomic \cite{allen1987ora} and a transmission line model with distributed components \cite{heaviside} as discussed for a general system of cascaded circuits \cite{0143-0807-26-6-019}. We briefly discuss this perspective in the appendix \ref{section_why}. This similarity interrogates the concept of propagation (slow or fast) in structured material which can be described by coupled propagation equations \cite{Winful82}. This approach serves as a basis for the theoretical analysis of Boyd \cite{Boyd:11} by distinguishing material and structural slow-light. Giving a unified vision of the different practical situations, distributed components, structured optical materials and atomic media is beyond the scope of our paper but definitely deserves further consideration.

While our goal in this work is to highlight the convergence of two domains exploiting a formal analogy between RLC resonators and two-level atoms, our approach can certainly be useful in different contexts. The recent possibilities offered by metamaterials \cite{PhysRevLett.101.253903, PhysRevB.85.073102, Gu:2012:10.1038/ncomms2153, PhysRevB.87.161110}, plasmonic structures \cite{taubert2012classical,Taubert:13}, optomechanical resonators \cite{PhysRevA.93.023802}, micro-resonators \cite{PhysRevLett.98.213904,PhysRevA.75.063833, Wang:16} and superconducting circuits \cite{PhysRevA.77.013831,  PhysRevLett.109.253603} to design at will the interaction with the electromagnetic field can certainly be guided by the analogy between RLC circuits and atoms. This complementarity can hopefully open perspectives in applied domains.
 
\section{Funding Information}

Agence Nationale de la Recherche DISCRYS (ANR-14-CE26-0037),

We would like to acknowledge George Heinze and Renaud Mathevet for fruitful discussions, Rose Ahlefeldt for technical assistance and the colleagues who encouraged us to write this paper.

\bibliographystyle{plain}

\appendix



\section{Telegrapher's equations} \label{section_why}

In the main part of the paper, we have modeled the series of discrete components and evaluated the transmission using transfer matrices.  We here use a description in terms of continuous functions as in the telegrapher's equations.

The reader who is familiar with the telegrapher's equation may be confused by our approach. This latter describes the ideally lossless propagation of signals in a transmission line. That is not what we aim at. On the contrary, we'd like to design an absorbing line with resonant losses as introduced in \ref{sec:ladder}. We simply generalized the telegrapher's equation with arbitrary impedances. A description of our topology in terms of propagation is important because slow-light is usually associated to a reduction of the group velocity $v_g=\displaystyle \frac{\partial \omega}{\partial k}$ in a medium whose response is characterized by a dispersion relation ($\omega$ is the excitation frequency and $k$ the wavevector). Our goal in this section is precisely to define a propagation constant and a wavevector.


\subsection{Transmission line model}\label{transmission_line}

%


As discussed in \ref{model} for the distributed components model, the most elementary object  is a two-port component as shown in fig.\ref{fig:bloc}. We keep the general case of $Z_0$ and $Z_1$ as complex impedances. Input/ouput relations for this single component that we wrote in a matrix form (eq.\ref{transfer_matrix}) are deduced from Kirchhoff's laws and read as:
\begin{eqnarray}\label{discrete}
\mathcal{V}_{n+1}-\mathcal{V}_{n} & = & Z_0 \, \mathcal{I}_{n} \nonumber \\
\mathcal{I}_{n+1}-\mathcal{I}_{n}& = & \frac{1}{Z_1}\, \mathcal{V}_{n+1}
\end{eqnarray}

These equations can be extended to continuous variables by writing $V$ and $I$ as functions of a propagation variable $x$.  The voltage and the current are then evaluated at a given position $x=n\times dx$ in the transmission line. In this equation, $dx$ is the characteristic length of the elementary component, which is introduced to give the propagation variable $x$ the dimension of a length. The total length of the line is then $X=N \times dx$. The correspondence between discrete and continuous functions is
\begin{eqnarray}\label{discrete2}
 V(n \,dx)=\mathcal{V}_n \nonumber \\
 I(n\, dx)=\mathcal{I}_n
\end{eqnarray}
While we have chosen to introduce $dx$, the elementary component length, we could equally use a dimensionless variable for $x$ to continuously describe the number of elements $n$, in which case $V(n)=\mathcal{V}_n$. The group velocity would then have the dimension of inverse time. This is a perfectly valid choice as discussed by Nakanishi \cite{nakanishi:323}. However, we prefer to introduce the unit cell length so that the group velocity will have its standard unit, length divided by time.

The equations \eqref{discrete} for discrete variables can be replaced by  their differential counterparts if the differential variation from one element to the next is small. This discretization condition can be written $\displaystyle \left| \x \right| \ll 1$. In what follows, we consider the situation where the discretization condition does apply, and eqs.\eqref{discrete} can be rewritten in their continuous form:

\begin{eqnarray}\label{diff_V_I}
\frac{\partial V(x)}{\partial x} & = & \frac{Z_0}{dx} \, I(x) \nonumber \\
\frac{\partial I(x)}{\partial x} & = & \frac{1}{Z_1 dx}\, V(x)
\end{eqnarray}

We will focus primarily on voltage measurements, which are described by the second order equation:
\begin{equation}\label{propag} 
\frac{\partial^2 V(x)}{\partial x^2}  =  \frac{Z_0}{Z_1 (dx)^2}\, V(x)
\end{equation}

Two boundary conditions are imposed to solve this equations. They are given on one side by the input voltage $\mathcal{V}_0$ and on the other side by the load impedance at the output.  As already mentioned, we choose a high load impedance, which means that $\mathcal{I}_N=0$. The general solution to eq.\eqref{propag} is then given by
\begin{equation} \label{eq:gensolution}
 V(x)  =    \mathcal{V}_0 \frac{e^{-\gamma x} +e^{\gamma x} e^{-2\gamma X} }{ 1+e^{-2\gamma X}}
\end{equation}
where we have introduced the propagation constant $\displaystyle \gamma = \frac{1}{dx}\sqrt{\x}$ whose imaginary part defines the absorption and the real part the dispersion. The two terms in \eqref{eq:gensolution} correspond to the forward and backward solutions in a broad sense.

The value of this function at the output of the transmission line (the transfer function of the line)  is

\begin{equation}\label{solution_V_L}
 V(X) =  \mathcal{V}_0 \frac{2 e^{-\gamma X}}{ 1+e^{-2\gamma X}} 
\end{equation}


The solution is completely general at the moment because there is no restriction on  $Z_0$ and $Z_1$ except the discretization condition; they are complex impedances, potentially including a frequency dependence  $\omega$. We now apply the general solution to the two extreme cases: a lossless transmission line (no resistance) as in the ideal telegrapher's equation the reader may be familiar with, and a purely resistive line, before moving on to the more complex situation required for slow-light demonstration. 

\subsubsection{Lossless transmission line with inductors and capacitors}\label{lossless}
A lossless transmission line is the situation considered in the classic telegrapher's equation. In this case, the impedances  $Z_0$ and $Z_1$ are purely imaginary, $Z_0=j L \omega$ and $Z_1=\displaystyle \frac{1}{j C \omega}$.  If the line is composed of discrete components, the continuous solution of eq.\eqref{eq:gensolution} can be applied if the discretisation condition $\displaystyle \left| \x \right| \ll 1$  is satisfied. In this system, this condition can be re-written $\omega \ll \displaystyle {1}/{\sqrt{LC}}$, where the quantity on the right side of this inequality is the cutoff frequency of the circuit \cite{0143-0807-26-6-019}. In this case, the propagation is described by a purely imaginary propagation constant,
\begin{equation}
\gamma=j \omega \sqrt{LC/(dx)^2}
\end{equation}
The term $LC/(dx)^2$ can be replaced by the product $\lambda_L \lambda_C$  of the distributed inductance $\lambda_L$ (henries per unit length)  and the capacitance $\lambda_C$ (farads per unit length).

Because $\gamma$ is imaginary, the solution to eq.\eqref{eq:gensolution} is composed of forward $+jkx$ and backward $-jkx$ propagating waves, where the wavevector  $k=\Im (\gamma) = \omega \sqrt{\lambda_L \lambda_C}$. The amplitudes of these waves are determined by the boundary conditions, and the backward propagating wave can be removed by choosing what is known as a matched load impedance. In our case we assume a high load impedance so there is no matching. This has the advantage of simplicity as $\mathcal{I}_N=0$ but it can be also difficult in practice when $Z_0$ and $Z_1$ have complex expressions. As we will see later, impedance matching is not necessary to mimic an absorbing medium.

\subsubsection{Purely resistive transmission line}\label{lossy}
The opposite case to a lossless transmission line is a purely resistive transmission line. We now consider the impedances $Z_0$ and $Z_1$  to be due to resistors  with $Z_0=R_0$ and $Z_1=R_1$.  The propagation constant $ \gamma$ is purely real, resulting in absorption, with an absorption coefficient $\alpha=\displaystyle \Re (\gamma) = \frac{1}{dx}\sqrt{\frac{R_0}{R_1}}$. In this situation, strictly speaking there is no propagation because $k=0$. The propagation constant $\gamma(\omega)$ is still well defined and the term  propagation should be understood in the general sense of an input/output relation.

The discretization condition can be written in terms of this absorption coefficient as $\alpha\, dx \ll 1$: the absorption per elementary cell is low. When this condition holds, the output voltage is given by
\begin{equation}\label{line_resistive}
V(X) =  \mathcal{V}_0 \displaystyle \frac{2 e^{-\alpha X}}{ 1+e^{-2\alpha X}} 
\end{equation}

In the strong absorption limit, when the total absorption $\alpha\, X$ is large, this can be approximated to zeroth order to yield a single exponential decay:
\begin{equation}\label{exp_dec}
V(X) =  {2 \mathcal{V}_0} e^{-\alpha X}
\end{equation}

Again the output of the transmission line depends on the boundary conditions, which can be modified by adjusting the load impedance. For a strongly absorbing medium, impedance matching is less critical because the backward term $-\alpha x$ is small. 

%
%
%
%
By requiring that the absorption per cell in the circuit is small while the total absorption is large, we have retrieved in  eq.\eqref{exp_dec} an equivalent of the Bouguer-Beer-Lambert attenuation law \cite{bouguer1760traite}. This is what we also predict using a discrete components model as discussed in \ref{abs_spectrum} (see the exponential law in fig. \ref{fig:Abs_delay_vs_Nelt}.c)

However, while we reproduce an exponential decay (Bouguer-Beer-Lambert law) using a transmission line model, the analogy with the Maxwell propagation equations in a dielectric medium is incomplete. In an optical medium, the electric field is described in the slowly varying envelope approximation by a first order differential equation \cite{allen1987ora} as opposed to  second order as in eq.\eqref{propag}. This leads to a single exponential decay and not a composition of $e^{-\alpha x}$ and $e^{\alpha x}$. This profound difference in terms of master equation and boundary conditions should be kept in mind when the analogy between a modified transmission line and an optically absorbing medium is considered.

For a reduction of the group velocity, absorption only is not sufficient. We now discuss how to produce a strong dispersive part.

%
%
%

\subsection{Introducing a non trivial dispersion}\label{introducing_TW}
The two situations presented above can be used to better understand the conditions required for slow-light. As stated earlier, slow-light $v_g \ll v_\phi$ requires  large dispersion.  Dispersion is given by the frequency dependence of the imaginary part of the propagation constant $\gamma$, which is precisely the wavevector $k$, while absorption is given by the real part of $\gamma$.

In Section \ref{lossless}, we considered a lossless transmission line, for which the propagation constant was purely imaginary,
$\gamma(\omega)=j \omega \sqrt{\lambda_L \lambda_C}$. This gives a phase velocity
\begin{equation}
v_\phi=\displaystyle \frac{\omega}{ k} = \frac{1}{ \sqrt{\lambda_L \lambda_C}}
\end{equation}
which has no frequency dependence. Therefore, there is no dispersion, and the group velocity is $v_g=\displaystyle \frac{\partial \omega}{\partial k}=v_\phi$.

In Section \ref{lossy}, we considered a resistive transmission line, which has $\gamma$ real. This situation displays absorption  with an absorption constant $\alpha= \displaystyle \frac{1}{dx}\sqrt{\frac{R_0}{R_1}}$. Again, there is no frequency dependence in this situation and the transmission line displays no dispersion. However, this strong absorption limit is interesting because of the link between the dispersion and absorption given by the well-known Kramers-Kronig relations. This implies that if the strong absorption seen in the resistive line is made frequency-dependent, a strong dispersion will be obtained, and slow-light propagation could be observed.

In the following section, we first consider the dispersive properties of a single absorption peak in a circuit (single resonant cell). Then, we address the more complex situation of two neighbouring peaks corresponding to the experimental situation presented in \ref{cell}.

\subsubsection{Dispersion of a single resonance cell}\label{dispersion_peak}

To generate an absorption peak rather than a flat absorption as in the resistive transmission line, we replace the resistive impedance $Z_1$ by a high-Q RLC circuit and keep $Z_0$ as purely resistive. With $Z_0=R_0$ and $Z_1(\omega)=\displaystyle R_1+jL\omega+\frac{1}{jC\omega}$, the solution in eq.\eqref{solution_V_L} has the same general form with: 
\begin{equation}
 \gamma(\omega) = \frac{1}{dx}\sqrt{\frac{Z_0}{Z_1(\omega)}}
 \end{equation}
 We can define for this system the resonant  frequency $\omega_0=\displaystyle \frac{1}{\sqrt{LC}}$ and the linewidth $\Gamma=\displaystyle \frac{R_1}{L}$. In the high-Q limit $\Gamma\ll \omega_0$ and close to resonance $\left|\omega-\omega_0\right|<\Gamma$), we can approximate the solution to first order to obtain for the resonance
\begin{equation}\label{alpha_1_RLC}
\displaystyle \gamma(\Delta) =  \alpha(\Delta)+i k(\Delta) = \frac{\alpha_0} {1+\displaystyle j\frac{\Delta}{\Gamma}}
\end{equation}
where $\Delta=\omega-\omega_0$ is the detuning from the resonance frequency and $\alpha_0= \displaystyle  \frac{1}{dx}\sqrt{\frac{R_0}{R_1}} $ the on-resonance absorption coefficient. In this equation,  we have retrieved an approximated Lorentzian frequency response analogous to that of a two-level atomic system.

Above, we chose $Z_1$ to be an RLC resonator because we know that this type of circuit has a frequency response similar to that of the two-level atoms that we are trying to model. However, we could have taken a different approach to design a circuit with the Lorentzian frequency response of eq.\eqref{alpha_1_RLC}. Electronics engineers have an approach based on network synthesis filters\cite{zverev1967handbook} that starts with the desired frequency response of eq.\eqref{alpha_1_RLC} and designs a circuit that creates that response. This approach would be very useful for modelling more complicated behaviors of atomic systems, where the circuit analogue is not immediately clear.

The rest of this section is essentially pedagogical. We define the group velocity and group delay from the dispersion relation. Starting from the propagation constant $\gamma$ (eq.\ref{alpha_1_RLC}), we can now derive the dispersion relation for this circuit. This is given by the wavevector $k(\Delta)$, which is now dependent on the detuning from the circuit resonance frequency\footnote{Instead of the wavevector $k(\omega)$, one can alternatively use the refractive index $n_r(\omega)=\displaystyle \frac{c}{v_\phi}=c\displaystyle \frac{k(\omega)}{\omega}$ to define the group velocity $v_g=\displaystyle \frac{c}{n_r+\omega \displaystyle \frac{\partial n_r}{\partial \omega}}$}:  
\begin{equation}\label{k_1_RLC}
\displaystyle k\left(\Delta\right) =-\alpha_0\frac{\Gamma \Delta} { \Delta^2+\Gamma^2}
\end{equation}
which allows us to define the group velocity
\begin{equation}\label{v_g_1_RLC}
\displaystyle v_g \left(\Delta\right)=\displaystyle \left(\displaystyle\frac{\partial k}{\partial \Delta}\right)^{-1} =\frac{\Gamma}{\alpha_0}\, \frac{\left(\Delta^2+\Gamma^2\right)^2} {\Gamma^2\left( \Delta^2-\Gamma^2\right)}
\end{equation}
In this equation, we retrieve the expression of the group velocity in a resonant dielectric medium, such as a two-level atomic ensemble \cite{brillouin1960wave}.
%
%

 At resonance, the group velocity is negative, giving rise to fast-light propagation. This is an interesting phenomenon, but it is often very difficult to explore in experimental atomic systems since the absorption is very high on resonance \cite{PhysRevLett.48.738}. In the wings of the resonance, where $|\Delta|>|\Gamma|$, the description by a Lorentzian is inaccurate, but we still expect the group velocity \eqref{v_g_1_RLC} to be positive leading to slow-light propagation.

Rather than use the group velocity to characterise the temporal response of the circuit in the experiment, we will use the group delay, which is more experimentally accessible. This is defined as $T_g^0=\displaystyle \frac{L}{v_g}$
\begin{equation}\label{T_g_1_RLC}
\displaystyle T_g^0 \left(\Delta\right) =\frac{\alpha_0 L}{\Gamma}\, \frac{\Gamma^2\left( \Delta^2-\Gamma^2\right)}{\left(\Delta^2+\Gamma^2\right)^2} 
\end{equation}

\subsubsection{Dispersion in a double resonance cell}\label{disptransparency}

In our circuit, the two resonances arise from two RLC circuits placed in parallel with well-resolved natural frequencies $\omega_a$ and $\omega_b$ (see fig. \ref{fig:ladder}). Guided by the apparent resemblance of the absorption curve (fig. \ref{Spectro_CW}) with two adjacent Lorentzian, we propose to write the propagation constant of the combined system as the sum
\begin{equation}\label{alpha_2_RLC}\begin{split}
\displaystyle \gamma(\omega)=  \alpha(\omega)+i k(\omega) = \frac{\alpha_0} {1+\displaystyle j\frac{\omega-\omega_a}{\Gamma}}+\\ \frac{\alpha_0} {1+\displaystyle j\frac{\omega-\omega_b}{\Gamma}}\end{split}
\end{equation}

following the discussion and the notations of \ref{dispersion_peak}.


Midway between the two resonances, the spectrum is symmetric and the detunings of both resonators are the same. This region is called a transparency window to take the slow-light propagation terminology. The dispersion of both resonators is the same and the group delay is the sum of the group delays from each resonance:
\begin{equation}
T_g= 2 T_g^0 \left(\delta\right) \longrightarrow {\alpha_0 L}\, \frac{\Gamma}{\Delta^2} \mbox{ for } {\delta \gg \Gamma}  
\end{equation}
with $\delta = | \omega_a - \omega_b|/2$ the half width of the transparency window. Having discussed an approximate expression for the dispersion relation, we know come back to the transfer function which is measured experimentally.

%

\subsubsection{Strong total absorption limit}\label{strong_total_abs}
We here briefly discuss the condition in which the transfer function is properly described by a single exponential decay (as in eq.\ref{exp_dec}) even when the propagation constant $ \gamma(\omega)$ is frequency dependent.

As stated in Section \ref{lossy}, the high absorption situation also supresses the backward-travelling part of the transfer function in eq.\eqref{solution_V_L}. So if $\alpha(\omega) X$ is large in the spectral region of interest, we can write the transfer function of the system of cascaded two-element RLC circuits as a single exponential decay:
\begin{equation}\label{forward_gamma}
V(X) =  {2 \mathcal{V}_0} e^{-\gamma(\omega) X}
\end{equation}

More specifically even in the transparency window between the two resonances, the absorption $\alpha(\omega) X$ is kept large so the expression \eqref{forward_gamma} is valid.  The term {\it transparency window} is still applicable, however, as the absorption in this region is much smaller than the resonant absorption of the two peaks.

An alternative option for eliminating the  backward wave would be to use the load impedance matching condition in the transparency window. This is  possible within the type of circuit used here, but is beyond the scope of the paper\footnote{The backward wave is eliminated for a load impedance $\sqrt{Z_0 Z_1(\omega)}$. For a lossless transmission line, this quantity, $\sqrt{\lambda_L/\lambda_C}$ is real and can be matched with an appropriate resistor, 50$\Omega$ or 75$\Omega$ depending on the line. In our case, $\sqrt{Z_0 Z_1(\omega)}$ keeps a dependence with $\omega$ and cannot be purely resistive. The load impedance requires a specific design.}. 

Our derivation is approximate at many levels. The description of the propagation constant as a Lorentzian is only to valid close to resonance (see \ref{dispersion_peak}), we simply extend it to obtain a simple analytical propagation formula (eq. \ref{forward_gamma}) also guided by the phenomenological resemblance of the absorption spectrum with Lorentzian curves in fig. \ref{Spectro_CW} (top).

As a validation a posteriori, we propose to use the fitted values of the components obtained in fig.\ref{Spectro_CW} (namely $R_0^\mathrm{fit}=4.7 \Omega$, $R_1^\mathrm{fit}=6.7 \Omega$, $L^\mathrm{fit}=25 \mu $H, $C_a^\mathrm{fit}=570 $ pF and $C_b^\mathrm{fit}=910 $ pF) to derive the propagation constant \eqref{alpha_2_RLC} and finally obtain the total transfer function as \eqref{forward_gamma}. It should be noted that the discretisation condition $\displaystyle \left| \x \right| \ll 1$  is poorly satisfied on resonance with $R_0^\mathrm{fit}=4.7 \Omega$ and $R_1^\mathrm{fit}=6.7 \Omega$. The result is plotted as a green line in fig. \ref{Spectro_CW}. The agreement is surprisingly correct even if the transmission line model only crudely apply.

This approach has the advantage to unambiguously define the propagation constant thus reinforcing the analogy between slow-light propagation and a ladder of RLC circuits.

\end{document}